 \documentstyle[12pt]{article}
\begin{document}
 \newcommand{\be}[1]{\begin{equation}\label{#1}}
 \newcommand{\ee}{\end{equation}}
 \newcommand{\beqn}[1]{\begin{eqnarray}\label{#1}}
 \newcommand{\eeqn}{\end{eqnarray}}
\newcommand{\mat}[4]{\left(\begin{array}{cc}{#1}&{#2}\\{#3}&{#4}\end{array}
\right)}
 \newcommand{\matr}[9]{\left(\begin{array}{ccc}{#1}&{#2}&{#3}\\{#4}&{#5}&{#6}\\
{#7}&{#8}&{#9}\end{array}\right)}
 \newcommand{\eps}{\varepsilon}
 \newcommand{\Ga}{\Gamma}
 \newcommand{\la}{\lambda}
\newcommand{\ov}{\overline}
\renewcommand{\thefootnote}{\fnsymbol{footnote}}

\begin{titlepage}
\begin{flushright}
INFN-FE-05-95 \\
hep-ph/9505384 \\
April  1995
\end{flushright}
\vspace{10mm}

 \begin{center}
 {\large\bf Predictive SUSY SO(10) Model with Very Low tan${\bf \beta}$ }

\vspace{5mm}
{\large  Zurab G. Berezhiani}
\footnote{E-mail: BEREZHIANI@FERRARA.INFN.IT,
31801::BEREZHIANI}\\ [5mm]
{\em Istituto Nazionale di Fisica Nucleare, Sezione di Ferrara, 44100 Ferrara,
Italy, \\ [2mm]
and\\ [2mm]
Institute of Physics, Georgian Academy
of Sciences, 380077 Tbilisi, Georgia}\\ [8mm]
\end{center}

\vspace{2mm}
\begin{abstract}


The first fermion family might play a key role in understanding the
structure of flavour: a role of the mass unification point.
The GUT scale running masses $\bar{m}_{e,u,d}$ are rather close,
which may indicate an approximate symmetry limit. Following this
observation, we present a new
predictive approach based on the SUSY $SO(10)$ theory
with $\tan\beta\sim 1$. The inter-family hierarchy is first generated
in a sector of hypothetical superheavy fermions and then transfered
inversely to ordinary
quarks and leptons by means of the universal seesaw mechanism.
The Yukawa matrices are simply parametrized by the
small complex coefficients $\eps_{u,d,e}$ which are related
by the $SO(10)$ symmetry properties.
Their values are determined by the ratio of the GUT
scale $M_X\simeq 10^{16}$ GeV to a higher (possibly string) scale
$M\simeq 10^{17}-10^{18}$ GeV.
The suggested ansatz correctly reproduces the fermion mass
and mixing pattern.
By taking as input the masses of leptons and $c$ and $b$ quarks,
the ratio $m_s/m_d$ and the value of the Cabibbo angle, the $u,d,s$ quark
masses, top mass and $\tan\beta$ are computed.
The top quark is naturally in the 100 GeV range, but with upper limit
$M_t<165$ GeV, while the lower bound $M_t>160$ GeV implies $m_s/m_d>22$.
$\tan\beta$ can vary from 1.4 to 1.7.
The proton decaying $d=5$ operators $qqql$ are naturally suppressed.

\end{abstract}

\end{titlepage}

\renewcommand{\thefootnote}{\arabic{footnote})}
\setcounter{footnote}{0}

\newpage

{\large \bf 1. Introduction }
\vspace{2mm}

Understanding the fermion mass spectrum is one of the main issues
in particle physics. In the standard model (SM) the Yukawa coupling
constants are arbitrary,
so one has to think of a more fundamental theory occuring at higher
energies. One of the most promising ways beyond the SM is related to
supersymmetric grand unified theories (SUSY GUTs) which provide a sound
basis for solving the issues of the gauge coupling unification and the
weak scale hierarchy. On the other hand, grand unification can also
play an important role in understanding the flavour structure,
by imposing specific constraints on the fermion mass matrices and
thus reducing the number of free parameters.
The $SO(10)$ GUT is a very appealing candidate for this purpose.
It unifies all quark and lepton states of one family into the
irreducible representation 16, providing thereby a possibility to link
their masses with certain group-theoretical relations.

In order to understand how the fermion mass spectrum could reflect
the grand unification features, it is necessary to compare the quark and
lepton running masses $\bar m_f$ or their Yukawa constants $\lambda_f$
($f=u,d,e,\dots$) at the GUT scale $M_X\simeq 10^{16}\,$GeV.
In the minimal supersymmetric standard model (MSSM)
these are related as $\bar{m}_f=\lambda_f v\sin\beta(\cos\beta)$, where
$v$ is the electroweak scale, $\sin\beta$ stands for the case of upper
quarks and $\cos\beta$ for the down quarks and charged leptons.
One can observe that the {\em vertical} mass splitting is small within
the first family and is quickly growing with the family number:
\be{vert}
\frac{\bar m_u}{\bar m_{d,e}}\sim 1\,,~~~~
\frac{\bar m_c}{\bar m_{s,\mu}}\sim 10\,,~~~~
\frac{\bar m_t}{\bar m_{b,\tau}}\sim 10^2
\ee
whereas the splitting between the charged leptons and down quarks
remains considerably smaller -- the third family is almost unsplit:
$\bar m_b\approx \bar m_\tau$, whereas the first two families are split
by a factor of about 3 but $\bar m_d \bar m_s\simeq \bar m_e \bar m_\mu$.

The {\em horizontal} hierarchy of quarks exhibits the approximate scaling low
\be{qh}
\frac{1}{\bar m_u}:\frac{1}{\bar m_c}:\frac{1}{\bar m_t}\sim
1:\eps_u:\eps_u^2\,,~~~~~
\frac{1}{\bar m_d}:\frac{1}{\bar m_s}:\frac{1}{\bar m_b}\sim
1:\eps_d:\eps_d^2\,
\ee
with $\eps_u^{-1}=200-300$ and $\eps_d^{-1}=20-30$, while for the charged
leptons we have
\be{lh}
\frac{1}{\bar m_e}:\frac{1}{\bar m_\mu}:\frac{1}{\bar m_\tau}
\sim 1:\eps'_e:\eps'_e\eps_e
\ee
with $\eps_e\sim \eps_d$ and $\eps'_e \sim \eps_u$.
In addition, the quark mixing angles have the following pattern:
\be{CKM}
s_{12}\sim \eps_d^{1/2}\,,~~~~ s_{23}\sim \eps_d\,,~~~~
s_{13}\sim \eps_d^{2}
\ee
Moreover, there are intriguing relations between
fermion masses and mixing angles, like the
well-known formula $\,s_{12}=(m_d/m_s)^{1/2}$ for the Cabibbo angle.

A popular idea is that the flavour structure is related
to the certain restricted form of mass matrices (e.g.
so called {\em zero textures} of refs. \cite{Fritzsch}), which can be
motivated by specific horizontal symmetry between fermion families.
Recently \cite{DHR} various zero texture ansatzes have been considered
on the basis of SUSY $SO(10)$ model and several interesting (and testable)
predictions were obtained for the fermion masses and mixing angles.
The key feature of this approach is that
the mass generation starts from the third family and proceeds to
the lighter ones through the smaller entries in the mass matrices
(in fact, this feature is generic for all models \cite{Fritzsch}).
Namely, the third family is directly coupled to the Higgs 10-plet,
so that $\lambda_{t,b,\tau}$ are equal at the GUT scale.
As for the lighter families, their masses are induced by certain
higher order operators with the specific $SO(10)$ structures.
The large splitting of the top and bottom masses  can be reconciled
only at the price of extremely large tan$\beta$ of about two orders of
magnitude, which can be achieved by certain tuning of parameters in the
Higgs potential \cite{tanb}. However, then it becomes rather surprising
that despite the giant tan$\beta$, $\bar m_c/\bar m_s$ is about 10 times
less as compared to $\bar m_t/\bar m_b$ while $\bar m_u$ and
$\bar m_d$ are almost unsplit. In order to achieve this, a judicious
selection of the $SO(10)$ Clebsch coefficients is required \cite{DHR}.

In the present paper we suggest an alternative approach to fermion
masses in a SUSY $SO(10)$ model with $\tan\beta\sim 1$. We follow the
observation that the masses of the first family exhibit an
{\em approximate} symmetry limit $\bar m_e\sim \bar m_u\sim \bar m_d$
(with splitting of about a factor of 2),
while the heavier families strongly violate it.
In the context of small tan$\beta$ this may indicate that the
$SO(10)$ Yukawa unification holds for the constants $\lambda_{u,d,e}$
rather than for $\lambda_{t,b,\tau}$.
How one could realize such a situation?

Nowadays the idea \cite{heavyferm} becomes popular that quark and
lepton masses are induced by the {\em universal seesaw} mixing with
hypothetical superheavy fermions, just in analogy with the famous seesaw
scenario for neutrinos. These are fermion states which have large
invariant mass terms or acquire masses after GUT breaking.
Thus, their exchanges induce the higher order effective operators
cutoff by the scale which can range from the Planck mass to the GUT scale.
With such a picture in mind, it is suggestive to think that $e,u,d$ are
`unsplit' since their masses are linked to an energy scale $M_1>M_X$
at which $SO(10)$ symmetry is still good, while the second and third
families are split being related to the lower scales
$M_{2,3}\leq M_X$ at which $SO(10)$ is no longer as good.

In particular, we assume that at the GUT scale the inverse Yukawa
matrices have the following form which we call the {\em inverse hierarchy
ansatz}:\footnote{ This pattern was first suggested in \cite{Rattazzi}
in the context of the radiative mass generation scenario. }
\be{mf-1}
\hat{\lambda}_{f}^{-1}= \frac{1}{\lambda}\,
(\hat{P}_1+\eps_f\hat{P}_2+\eps_f^2\hat{P}_3 )\,,~~~~~~~f=u,d,e
\ee
where the small complex expansion parameters
$\eps_f=\eps_u,\eps_d,\eps_e$ are different for the upper quark,
down quark and lepton mass matrices, and $\hat{P}_{1,2,3}$ are
some symmetric rank-1 matrices with $O(1)$ elements.
Without loss of generality, their basis can be chosen as
\be{Pbasis}
\hat{P}_1\!=\!(1,0,0)^T\!\!\bullet(1,0,0), ~~~
\hat{P}_2\!=\!(a,b,0)^T\!\!\bullet(a,b,0), ~~~
\hat{P}_3\!=\!(x,y,z)^T\!\!\bullet(x,y,z)
\ee
so that the inverse Yukawa matrices are alligned and have the form
\be{invform}
\hat{\la}_f^{-1}\,=\frac{1}{\la}
\matr{1+a^2\eps_f+x^2\eps_f^2}{ab\eps_f+xy\eps_f^2}{xz\eps_f^2}
{ab\eps_f+xy\eps_f^2}{b^2\eps_f+y^2\eps_f^2}{yz\eps_f^2}
{xz\eps_f^2}{yz\eps_f^2}{z^2\eps_f^2} \approx
\frac{1}{\la}
\matr{1+a^2\eps_f}{ab\eps_f}{xz\eps_f^2}
{ab\eps_f}{b^2\eps_f}{yz\eps_f^2}
{xz\eps_f^2}{yz\eps_f^2}{z^2\eps_f^2}
\ee
In lowest order their eigenvalues are given by diagonal entries:
$\lambda_{fi}\approx\lambda\eps_f^{1-i}$ ($i=1,2,3$ is a family index).
In this way, the quark mass pattern (\ref{vert}),(\ref{qh}) is
understood by means of $\eps_u\ll\eps_d$: $~\lambda_{u,d}\sim\lambda$,
$~\lambda_c/\lambda_s \sim (\eps_d/\eps_u)\sim 10$ and
$\lambda_t/\lambda_b\sim (\eps_d/\eps_u)^2\sim 10^2$.
This also implies that the CKM mixing emerges dominantly from the down
quark matrix $\hat{\lambda}_d$:
$\hat{\lambda}_u$ is much more "stretched" and essentially
close to its diagonal form, so that it brings only $O(\eps_u/\eps_d)$
corrections to the CKM mixing angles. Thus, at lowest order one expects
that $s_{12},s_{23}\sim \eps_d$ and $s_{13}\sim\eps_d^2$, which
estimates are indeed good for $s_{23}$ and $s_{13}$. For example,
from (\ref{invform}) follows that
$s_{23}\simeq |yz/b^2|\eps_d$ and $\la_s/\la_b\simeq |z^2/b^2|\eps_d$.
Then the experimental observation $s_{23}\sim \la_s/\la_b\sim \eps_d$
implies that $z\sim y\sim b$, so that the $\eps_f^2y^2$ term
in (2,2) element can be safely neglected as compared to the leading
term $\eps_f b^2$. The similar consideration for $s_{13}$ ensures that
$O(\eps^2)$ terms are negligible also in (1,1) and (1,2) elements.

However, the naive picture with also $\eps_{d,e}a^2\ll 1$ and distinct
$\eps_{e,d}$ immediately encounters the following problems:
{\em (i)} the Yukawa couplings $\la_{e,u,d}$ remain unsplit,
{\em (ii)} $s_{12}\sim \eps_d$ is too small as compared
to the actual value of the Cabibbo angle ($\sim \sqrt{\eps_d}$),
{\em (iii)} leptons behave as $\lambda_e : \lambda_\mu : \lambda_\tau
\sim 1 : \eps_e^{-1} : \eps_e^{-2}$, in contradiction to (\ref{lh}),
{\em (iv)} the {\em grand prix}, $b-\tau$ unification is lost:
$\la_b$ and $\la_\tau$ emerge at $O(\eps^2)$ level and e.g. the factor
of 2 difference among $\eps_{d}$ and $\eps_{e}$ would cause already
factor of 4 splitting between $\lambda_b$ and $\lambda_\tau$.

In this paper we show that all these problems can be naturally
solved in the framework of SUSY $SO(10)$ model.
As it was argued in \cite{Rattazzi}, the experimental value of the
Cabibbo angle $s_{12}\simeq (m_d/m_s)^{1/2}$ implies that
$|\eps_da^2|\simeq 1$. On the other hand, $SO(10)$ symmetry provides the
specific relation $\eps_e= -\eps_d-2\eps_u$ (see below, eq. (\ref{eps2})),
which ensures that $\la_b\approx \la_\tau$ due to the large $\la_t-\la_b$
splitting ($\eps_u\ll\eps_d$). In addition, then $\lambda_d$ and
$\lambda_e$ can split from $\lambda_u\approx \lambda$ to different
sides by about a factor of 2.

The paper is organized as follows. In the next section we demonstrate
how the Yukawa matrices of the form (\ref{mf-1}) can be obtained in the
context of the SUSY $SO(10)$ model \cite{old},  and study
implications of our scheme for the fermion masses and mixing.
Section 4 is devoted to a brief discussion of our results.

\vspace{6mm}
{\large \bf 2. Inverse Hierarchy Picture in SUSY ${\bf SO(10)}$ Model}
\vspace{2mm}

Consider a SUSY $SO(10)$ model
with three light fermion families $16^f_i$ and three families of
superheavy fermions $16^F_i+\overline{16}^F_i$, $i=1,2,3$.
It is convenient to describe the field content in terms of the
Pati-Salam $G_{PS}=SU(4)\otimes SU(2)_w\otimes SU(2)$ subgroup of $SO(10)$:
\beqn{16f}
16^f_i\!=\!f_i(4,2,1)+f^c_i(\bar{4},1,2),~~~~~~~~~~~~~~~~~~~~~\nonumber \\
16^F_i\!=\!{\cal F}_i(4,2,1)+F^c_i(\bar{4},1,2),~~~~
\overline{16}^F_i\!=\!{\cal F}^c_i(\bar{4},2,1)+F_i(4,1,2)
\eeqn
(Notice that ${\cal F}$'s are weak isodoublets and $F$'s are isosinglets).
We introduce also the Higgs 45-plets (45=(15,1,1)+(1,3,1)+(1,1,3)+(6,2,2))
of the following three types: $45_{BL}$ with VEV $V_{BL}$ on the (15,1,1)
fragment,
$45_R$ with VEV $V_R$ on the (1,1,3) fragment,
and $45_X$ having VEV $V_X$ shared by both (15,1,1) and (1,1,3)
components.\footnote{ The VEV orientation of these 45-plets are
determined by their couplings to the Higgs superfields 54 and
$16_H+\ov{16}_H$, which are also needed for the $SO(10)$ symmetry
breaking down to $SU(3)\otimes SU(2)_w \otimes U(1)$
(see e.g. \cite{BabuBarr}). In particular, $45_X$ has VEVs towards both
(15,1,1) and (1,1,3) fragments if superpotential includes the
terms $45_X^2 54$ and $45_X 16_H \overline{16}_H$.
As for $45_{BL}$ and $45_R$, in order to ensure the strict `zeroes' in
their VEVs, they should couple only to 54 but not to $16_H,\ov{16}_H$.
The trilinear term $45_{BL} 45_R 45_X$ evades the
unwanted Goldstone modes. }
%

For the electroweak symmetry breaking and the quark and lepton mass
generation  we use a traditional Higgs supermultiplet
$10=\phi(1,2,2)+T(6,1,1)$.
In order to maintain the gauge coupling unification, we
assume that all VEVs $V_{BL}$, $V_R$ and $V_X$ are of the order of
$M_X\simeq 10^{16}\,$GeV, and below this scale SUSY $SO(10)$ theory
reduces to the MSSM with three fermion families $f_i$ and
a couple of the standard Higgs doublets $h_{1,2}$ contained in $\phi$.
The field $45_{BL}$ serves for the solution of the doublet-triplet
splitting problem through the ``missing VEV'' mechanism \cite{DiWi}.
In this way the Higgs doublets $h_{1,2}$ are kept light
while their colour triplet partners contained in $T(6,1,1)$ acquire the
$O(M_X)$ mass (otherwise they would cause unacceptably fast proton decay
and also would affect the unification of gauge couplings).
The VEVs of $h_{1,2}$ arise radiatively after the SUSY
breaking:
\be{phi}
\langle\phi\rangle=\mat{v_2}{0}{0}{v_1}; ~~~~~~
(v_1^2+v_2^2)^{1/2}=v=174\,\mbox{GeV},~~~~\frac{v_2}{v_1}=\tan\beta
\ee

Let us assume that the direct Yukawa couplings $16^f16^f 10$ are
forbidden by certain symmetry reasons and the $16^f$'s get mass through
the `seesaw' mixing \cite{heavyferm} with their heavy partners
$16_F+\ov{16}_F$. The relevant terms in the superpotential are chosen as
\footnote{ We do not specify the symmetries leading to this pattern,
which question deserves special consideration.
The higher order operators cutoff by scale $M$ can be
induced by the exchanges of heavy (with masses $\sim M$)
fermion superfields in $16+\ov{16}$ representations,
so that the combinations in brackets transform as
effective 16 or $\overline{16}$.  }
\beqn{WY}
W_{fF}=\Gamma_{ij} 10\,16^f_i 16^F_j +
\frac{G_{ij}}{M}\,(\overline{16}^F_i 45_R)(45_R 16^f_j)
{}~~~~~~~~~~~~~~~~\nonumber \\
W_F= M Q_1^{ij} 16^F_i \overline{16}^F_j +
Q_2^{ij} 16^F_i 45_X \overline{16}^F_j +
\frac{Q_3^{ij}}{M}\,(16^F_i 45_X)(45_X\overline{16}^F_j)
\eeqn
where $M\gg M_X$ is some large (string?) scale, and $\hat{\Ga},
\hat{G}, \hat{Q}_{1,2,3}$
are the coupling constant matrices with $O(1)$ elements.
In what follows we do not specify any concrete texture, assuming
only that $\hat{\Ga}$ and $\hat{G}$ are arbitrary nondegenarate
matrices and $\hat{Q}_{1,2,3}$ are some rank-1 matrices.
After substituting large VEVs the whole $9\times 9$ Yukawa matrix
for the fermions of different charges gets the form
(each entry is $3\times 3$ matrix in itself):
\be{Mtot}
\begin{array}{ccc}
 & {\begin{array}{ccc} \,f^c & \,\,\,\;F^c & \,\,\;{\cal F}^c
\end{array}}\\ \vspace{2mm}
\begin{array}{c}
f \\ F \\ {\cal F}   \end{array}\!\!\!\!\!&{\left(\begin{array}{ccc}
0 & \hat{\Gamma}\phi  & 0 \\ \hat{M}_R  & \hat{M}_F & 0 \\
\hat{\Gamma}^{\dagger}\phi
 & 0 & \hat{M}_{\cal F} \end{array}\right)}
\end{array}, ~~~~~~~~ f=u,d,e,\nu
\ee
where the $(2,1)$-block
$\hat{M}_R=\eps_R^2\hat{G} M$ is the same for the fermions of all charges.
The $(1,2)$ blocks are also the same: the matrix $\hat{\Gamma}$
stands for the coupling of up-type and down-type fermions
with the MSSM Higgses $h_2$ and $h_1$, respectively.
The $(1,3)$-block is vanishing since the VEV
$\langle 45_R\rangle$ has the $(1,1,3)$ direction, so that the
${\cal F}$-type fermions are irrelevant for the seesaw mass
generation.\footnote{
This leads to natural suppression of the dangerous
$d=5$ operators inducing the proton decay:
since the $f$ and ${\cal F}$ states are unmixed,
the colour triplets in $T(6,1,1)$ can cause transitions of $f$'s
only into the superheavy ${\cal F}$'s. Thus, the $LLLL$-type
operators $[qqql]_F$ which bring the dominant contribution
to the proton decay
automatically vanish. As for the $RRRR$ type operators
$[u^cu^cd^ce^c]_F$, they occur due to the $f^c-F^c$ mixing
and have the usual strength. However, these are known to be
more safe \cite{Nath}. }
Thus, all information on the flavour structure is essentially
contained in the matrices of $F$ fermions
$\hat{M}_F=M\hat{Q}_F=M(\hat{Q}_1+\eps_f\hat{Q}_2+ \eps_f^2\hat{Q}_3)$,
where $\eps_f\sim V_X/M$ are the small, generally complex  parameters.
Since $45_X$ has VEVs both in (15,1,1) and (1,1,3) directions,
they have the form
$\eps_{d,u}=\eps_{15}\pm \eps_3$, $\eps_{e,\nu}=-3\eps_{15}\pm\eps_3$.
Therefore, only two of these four parameters are independent:
\be{eps2}
\eps_e=-\eps_d-2\eps_u\,,~~~~~~~~ \eps_\nu=2\eps_e+3\eps_u
\ee

After decoupling the heavy states in (\ref{Mtot}) our theory
reduces to the MSSM with the Yukawa coupling matrices
given by the following expression \cite{Rattazzi}:
\be{exactseesaw}
(\hat{\la}_f \hat{\la}_f^{\dagger})^{-1}\! =
(\hat{\Gamma}^{\dagger})^{-1}\left[1+\hat{M}_F^{\dagger}
(\hat{M}_R \hat{M}_R^{\dagger})^{-1} \hat{M}_F \right] \hat{\Gamma}^{-1}
\! = (\hat{\Gamma}\hat{\Gamma}^{\dagger})^{-1}
+ \frac{1}{\eps_R^4} (\hat{G}^{-1}\hat{Q}_F \hat{\Gamma}^{-1})^{\dagger}
(\hat{G}^{-1}\hat{Q}_F \hat{\Gamma}^{-1})
\ee
When $\hat{M}_R\gg\hat{M}_F$, this equation gives the obvious
result $\hat{\la}_f=\hat{\Gamma}$. On the other hand, for
$\hat{M}_R\ll \hat{M}_F$ it reduces to the ``seesaw'' formula
$\hat{\la}_f=\hat{\Gamma}\,\hat{M}_F^{-1}\,\hat{M}_R$, so that we have
\be{Yf-1}
\hat{\la}_{f}^{-1}=\frac{1}{\eps_R^2}\,\hat{G}^{-1}
\hat{Q}_F \hat{\Gamma}^{-1}=
\frac{1}{\la}(\hat{P}_1+\eps_f\hat{P}_2+\eps_f^2\hat{P}_3)\,,
\ee
where $\lambda=\frac{1}{N}\eps_R^2$ and
$\hat{P}_n=\frac{1}{N} \hat{G}^{-1}\hat{Q}_n \hat{\Gamma}^{-1}$
are still rank-1 matrices. For definiteness, the normalization factor
$N$ is chosen as the nonzero eigenvalue of the matrix
$\hat{G}^{-1}\hat{Q}_1\hat{\Gamma}^{-1}$.


The seesaw limit is certainly very good for all light states
apart from $t$ quark: their Yukawa couplings are much smaller than $1$,
so that the first term in (\ref{exactseesaw}) can be safely neglected.
However, as far as $\lambda_t\sim 1$ (or, in other words,
$\eps_R\sim \eps_u \ll \eps_{d,e}$), for its evaluation one has
to use the exact formula (\ref{exactseesaw}).
Then the top genuine constant $\tilde{\la}_t$ is related to the
`would-be' Yukawa coupling $\la_t$ of the seesaw limit (\ref{Yf-1}) as
\be{la-t}
\tilde{\la}_t=\frac{\la_t}{\sqrt{1+(\la_t/\Gamma_c)^2}}<\la_t
\ee
where  $\Gamma_c$ is a certain combination of the constants in
$\hat{\Gamma}$. For example, in the basis of $\hat{Q}_n$ having
a form similar to (\ref{Pbasis}), $\Gamma_c^2=\sum |\Gamma_{3i}|^2~$
($i=1,2,3$).

In what follows, we assume for simplicity that $\hat{P}_n$ are symmetric.
This will not change essentially our results (see comment
at the end of this section). Then the inverse Yukawa matrices
at the GUT scale have the form (\ref{invform}) which,
as was already noted, must be diagonalized by assuming
$a^2\eps_f\sim 1$. Thus, the Yukawa eigenvalues are
\be{eigen}
\la_{u,d,e}=\frac{\la}{|1+a^2\eps_{u,d,e}|}\,,~~~~~
\la_{c,s,\mu}=\frac{\la |1+a^2\eps_{u,d,e}|}{|b^2\eps_{u,d,e}|}\,,~~~~~
\la_{t,b,\tau}=\frac{\la}{|z^2\eps_{u,d,e}^2|}
\ee
and for the CKM angles we obtain
\be{Cabibbo}
s_{12}=\frac{|\eps_dab|}{|1+\eps_da^2|}=
\sqrt{\frac{\la_d}{\la_s}|\eps_da^2|}\,,~~~~
s_{23}=\frac{\lambda_u}{\lambda_d}\,\left|\frac{yz}{b^2}\eps_d\right|\,,~~~~
s_{13}=\frac{\lambda_d}{\lambda_u}\,|xz\eps_d^2|
\ee
These Yukawa constants are linked to the physical fermion masses through
the renormalization group (RG) equations. For the heavy quarks
$t,b,c$ we take their running masses at $\mu=m_{t,b,c}$,
while for the light quarks $u,d,s$ at $\mu=1\,$GeV. Then
we have (see e.g. \cite{RG}):
\beqn{RG}
m_u=\la_u\eta_u A_u B_t^3 v\sin \beta\,,~~~~
m_d=\la_d\eta_d A_d v\cos \beta\,,~~~~
m_e=\la_e\eta_e A_e v\cos \beta  \nonumber  \\
m_c=\la_c\eta_c A_u B_t^3 v\sin \beta\,,~~~~
m_s=\la_s\eta_d A_d v\cos \beta\,,~~~~
m_{\mu}=\la_{\mu}\eta_e A_e v\cos\beta \\
m_t=\tilde{\la}_t A_u B_t^6 v\sin \beta\,,~~~~
m_b=\la_b\eta_b A_d B_t v\cos \beta\,,~~~~
m_{\tau}=\la_{\tau}\eta_{\tau} A_e v\cos \beta \nonumber
\eeqn
where the factors $A_f$ account for the gauge coupling induced running
from the scale $M_X$ to the SUSY breaking scale $M_S\simeq m_t$,
the factors $\eta_f$ encapsulate the QCD+QED running from $M_S$ down to
$m_f$ (or to $\mu=1$ GeV for the light quarks $u,d,s$),
and $B_t$ includes the running induced by the top quark Yukawa coupling:
\be{B-t}
B_t=\exp\left[-\frac{1}{16\pi^2}\int_{\ln M_S}^{\ln M_X}
\tilde{\la}_t^2(\mu) {\mbox d}(\ln\mu)\right]
\ee

{}From (\ref{eigen}) we immediately obtain the relations
\be{bottom1}
\sqrt{\frac{\la_b}{\la_t}}=
\frac{\la_d\la_s}{\la_u\la_c}=\left|\frac{\eps_u}{\eps_d}\right|,
{}~~~~~~~~~~
\sqrt{\frac{\la_b}{\la_{\tau}}}=
\frac{\la_d\la_s}{\la_e\la_\mu}=\left|\frac{\eps_e}{\eps_d}\right|=
\left|1+2\,\frac{\eps_u}{\eps_d}\right|
\ee
When I get to the bottom I go back to the top:
as far as $|\eps_u/\eps_d|=\sqrt{\lambda_b/\lambda_t}\ll 1$, we have
the approximate relationships $\lambda_b\approx \lambda_\tau$ and
$\lambda_d\lambda_s\approx \lambda_e\lambda_\mu$.
Therefore, we are
not loosing the understanding of $b-\tau$ unification in spite of naive
expectation: it is more precise the more $t-b$ are split, and
this is granted by the $SO(10)$ symmetry relation (\ref{eps2}).

On the other hand, we know that $m_\mu/m_e=207$, $m_s/m_d\approx 22$
and $s^{-1}_{12}\approx \sqrt{22}$ -- {\em c'est la vie}.
Then (\ref{Cabibbo}) implies that $|\eps_da^2|\simeq 1$.
In addition, from $\la_d\la_s=\la_e\la_{\mu}$ follows
that $\la_d/\la_e=\la_\mu/\la_s=(m_d m_\mu/m_s m_e)^{1/2}\simeq 3$.
According to (\ref{eigen}), for $|\eps_{d,e}a^2|\simeq 1$ such a
splitting is indeed possible: owing to the relation
$\eps_d\approx -\eps_e$, $|1+\eps_da^2|$ and $|1+\eps_ea^2|$
can split to different sides from 1 by about a factor of 2.
Then the order of magnitude difference between
$\la_\mu/\la_e$ and $\la_s/\la_d$ can be naturally understood.
In this way, owing to the numerical coincidence
$(\la_e/\la_d)^2\!\sim\! \eps_u/\eps_d\sim 0.1$, we
reproduce the mixed behaviour (\ref{lh}) of leptons.
Thus, splitting of the first family and large value of the
Cabibbo angle (as compared to other mixing angles which have their
natural size $s_{23}\sim\eps_d$ and $s_{13}\sim \eps_d^2$)
have the same origin, namely
$|\eps_{d,e}a^2|\simeq 1$.\footnote{ One may question how to achieve
$\eps_da^2\simeq 1$, if the  Yukawa couplings are $O(1)$
and $\eps_d\sim 1/20-1/30\,$ (see (\ref{qh})). However,
in (\ref{Yf-1}) the Yukawa matrices $\hat{Q}_n$ are "sandwiched"
as $\hat{P}_n=\frac{1}{N}G^{-1}\hat{Q}_n \Gamma^{-1}$ so that $a,b\dots$
are actually given by the Yukawa constant ratios.
Thus, $a\sim 5$ could easily occur due to some
spread in the Yukawa couplings,
while the latter themselves are small enough to fulfill the
perturbativity bound $G^2_{Y}/4\pi<1$. As we noted earlier, the
mass and mixing pattern of the second and third families
suggests that such an accidental enhancement
does not happen for other entries in the matrix (\ref{invform}). }
Eqs. (\ref{eigen}) also imply
\be{u/d}
\frac{\la_u}{\la_d}=\frac{|1+\eps_da^2|}{|1+\eps_u a^2|}
{}~~~~~\Longrightarrow~~~~ \frac{\la_u}{\la_d}\leq
(1+|\eps_ea^2|)\left(\frac{m_em_s}{m_\mu m_d}\right)^{1/2}
\approx 0.6-0.7
\ee
so that the ratio $m_u/m_d\approx
(\la_u/\la_d)B_t^3\tan\beta~$ is less than 1 for small enough tan$\beta$.

The detailed numerical study of our ansatz leads to more concrete results.
We consider the masses of leptons and heavy quarks $c$ and $b$,
the ratio $\zeta=m_s/m_d$ and the Cabibbo angle $s_{12}$ as input,
and try to calculate other quantities.
For definiteness we take $\alpha_3(M_Z)=0.11$, $m_b=4.4$ GeV and
$m_c=1.32$ GeV, and use
for the RG running factors the results of ref. \cite{RG}.
Our computational strategy is the following:

$\bullet$ Substituting the lepton masses in
(\ref{RG}), we find $\la_e,~\la_\mu$ and $\la_\tau$ in terms of tan$\beta$.
Analogously, by fixing the values $m_c$ and $m_b$ we find
$\la_c$ and $\la_b$ in terms of $\tan\beta$ and $\tilde{\la}_t$.
In particular, we have $\lambda_\tau/\lambda_b=R_{b/\tau}B_t m_\tau/m_b$,
where $R_{b/\tau}=\eta_b A_d/\eta_\tau A_e\approx 3.1$ for
$\alpha_3(M_Z)=0.11$ \cite{RG}. Thus, $\lambda_b=\lambda_\tau$ is
achieved when $\lambda_t\simeq 1.5$ ($B_t\approx 0.8$).
Then by running the second equation (\ref{bottom1}) from the GUT scale
down to $\mu=1$ GeV, we readily obtain
\be{mdms}
m_dm_s = R_{d/e}^2 R_{b/\tau}^{-1/2} B_t^{-1/2}
m_e m_\mu \left(\frac{m_b}{m_\tau}\right)^{\frac{1}{2}} =
1100\,B_t^{-1/2}~ \mbox{MeV}^2
\ee
(notice the very weak dependence on $\tilde{\la}_t$),
so that for a fixed value of $\zeta$ we get
\be{md-ms}
m_d=7 \cdot(22/\zeta)^{1/2} B_t^{-1/4}\,\mbox{MeV}\,,~~~~~~~
m_s=155 \cdot (\zeta/22)^{1/2} B_t^{-1/4}\,\mbox{MeV}
\ee

$\bullet$ For $\Ga_{c}$ fixed, the first equation (\ref{bottom1})
determines $|\eps_u/\eps_d|$ as a function of $\tilde{\la}_t$ and
tan$\beta$. Then from the second equation also arg$(\eps_u/\eps_d)$
can be found in terms of $\tilde{\la}_t$ and tan$\beta$.

$\bullet$ The modulus $|\eps_da^2|$ is fixed by the value
of the Cabibbo angle:  $|\eps_da^2|\approx \zeta s_{12}^2$,
whereas $\arg(\eps_da^2)$ can be found from the equation
\be{d/e}
\frac{\la_d}{\la_e}=
\frac{|1-\eps_da^2(1+2\eps_u/\eps_d)|}{|1+\eps_da^2|}=
\left(\zeta\frac{m_e}{m_\mu}\right)^{-1/2}
\left(\frac{m_\tau}{m_b} R_{b/\tau} B_t \right)^{-1/4}
\ee

$\bullet$ In this way, the complex parameters $\eps_{u,d,e}a^2$ are
all expressed in terms of as yet unknown $\tan\beta$ and $\tilde{\la}_t$.
Then, using (\ref{u/d}) and (\ref{RG}) we find the mass ratio
$\rho=m_u/m_d$ as a function of $\tan\beta$ and $\tilde{\la}_t$.
The isocurves of $\rho$ are shown in Fig. 1 (dotted).

$\bullet$ Once $\eps_{e,u}^2a^2$ are known, from (\ref{eigen}) we find
$\la_u=\la_e|(1+\eps_ea^2)/(1+\eps_ua^2)|$ and substitute
it in the equation $\lambda_t/\lambda_\tau=
(\lambda_u\lambda_c/\lambda_e\lambda_\mu)^2$.
Then for fixed $\Gamma_{c}$ the latter becomes a relation which
determines tan$\beta$ as a function of $\tilde{\lambda}_t$
(see solid curves in Fig. 1).
We would like to stress that the chosen values of $\alpha_3(M_Z)$,
$m_b$ and $m_c$  are taken at their experimentally allowed
extremes.\footnote{
The controversy concerning the value of $\alpha_3(M_Z)$ is not
resolved yet. The $\Upsilon$ sum rules analysis implies
$\alpha_3(M_Z)=0.109 \pm 0.001$ \cite{Shifman}, whereas the SM global
fits based on the LEP/SLD data lead to $\alpha_3(M_Z)=0.127\pm 0.005$
\cite{Lang}. However, as it was argued in \cite{contra}, the
systematic error in the latter value, essentially
determined by analyzing $\Gamma(Z\rightarrow$ hadrons) can be largely
underestimated. The gauge coupling unification,
without taking into account the model-dependent threshold corrections,
also requires $\alpha_3(M_Z)>0.12$ \cite{Nir}.
In our $SO(10)$ model, however, $\alpha_3(M_Z)=0.11$ can be easily
adopted due to the threshold corrections emerging e.g. due to large
splitting in the third superheavy family $16_F + \ov{16}_F$.
In particular, the mass of the weak isosinglet upper quark of this
family is $\sim \eps_u^2 M$, which is about 2 orders of magnitude
below the masses of other fragments ($\sim \eps_d^2 M< M_X$). }
The value of tan$\beta$ decreases for smaller $m_b,m_c$ or larger
$\alpha_3(M_Z)$, so that the solid contours actually mark the upper
borders of allowed regions.

$\bullet$ Another relation between $\tilde{\la}_t$ and tan$\beta$
emerges by fixing the top pole mass  $M_t=m_t[1+4\alpha_3(m_t)/3\pi]$
(see dashed curves in Fig. 1).
The flat behaviour of these curves for large $\tilde{\la}_t$ corresponds
to infrared fixed regime when $\tilde{\la}_tB_t^6$ is practically
independent of $\tilde{\la}_t$ and the top pole mass is essentially
determined by tan$\beta$: $M_t=\sin\beta\cdot 190\,$GeV \cite{inffix}.

The results of numerical computations
are shown in Fig. 1. We see that the constant $\Ga_{c}$ which sets the
seesaw `cutoff' (\ref{la-t}) should be quite close to the
perturbativity bound in order to ensure the sufficiently large $M_t$.
Fig. 1 shows that at the perturbativity border ($\Ga_c=3.3$),
$M_t$ reaches the maximum when $\tilde{\la}_t\simeq 1.5$,
close to the infrared fix-point \cite{inffix} (we remind that in this
case $\la_b=\la_\tau$).
Namely, for $\Ga_{c}=3.3$ and $\zeta=22$
the maximal top mass is $M^{\rm max}_t=160$ GeV,\footnote{ By relaxing
the perturbativity bound, $M^{\rm max}_t$ can increase only by few GeV's:
see  solid curve for $\Ga_{c}=6.6$ in Fig. 1A.
The flat behaviour for large $\tilde{\la}_t$ can be easily understood.
According to (\ref{la-t}), for $\Ga_{c}\gg 1$ we have
$\tilde{\la}_t=\la_t$, while $\la_c\propto m_c/B_t^3$ and the ratio
$\la_u/\la_e\simeq 2$ very  weakly depends on $\la_t$. Then the
equation $\la_t/\la_\tau=(\lambda_u\lambda_c/\lambda_e\lambda_\mu)^2$
in fact determines the combination
$\la_t B_t^6 \tan^2\beta\cos\beta \propto M_t \tan\beta$
which is practically independent of $\la_t$ for enough large values
of the latter.  }
which corresponds to $\tan\beta=1.5$ and  $m_u/m_d=0.5-0.6$.
The latter values combined with $m_s/m_d=22$ perfectly fit the famous
$\rho-\zeta$ ellipse \cite{uds}. For smaller $\Ga_c$ or smaller $\zeta$
the $M^{\rm max}_t$ sharply decreases
(e.g. for $\zeta=19$ the maximal top mass can be at most 150 GeV).
On the other hand, for $\zeta=25$ we obtain $M^{\rm max}_t=165$ GeV,
which corresponds to $\tan\beta=1.7$ and $m_u/m_d= 0.6-0.8$ (see Fig. 1B).
The latter values seem too large versus $m_s/m_d=25$ \cite{uds}.

Taking all these into account, we see that the preferable
choice of the parameter region corresponds to $m_s/m_d\approx 22$,
when $m_u/m_d=0.5-0.6$ and $M_t\approx 160$ GeV ($\tan\beta\approx 1.5$),
at the lower edge of the recent CDF result
$M_t=176\pm 8\pm 10$ GeV \cite{CDF}.
By taking all input parameters at their extremes and also neglecting
the perturbativity constraint,
the maximal value of $M_t$ in our model can be increased at most up
to $170$ GeV.

Let us conclude this section with the following comment.
The symmetric form of the matrices $\hat{P}_n$ in (\ref{Yf-1}) was
imposed by hands. In fact, it can be ensured by introducing certain
horizontal symmetries, or
by extending the gauge symmetry e.g. to $SO(10)\otimes SO(12)$,
with $16_f$'s belonging to $SO(10)$ and $16_F+\ov{16}_F$'s contained
in  32-plets of $SO(12)$ \cite{inprep}. The latter case implies
that $\hat{Q}_n$ are symmetric and $\hat{G}=\Ga^T$.
On the other hand, for non-symmetric $\hat{P}_n$ instead of
(\ref{Cabibbo}) we obtain the equation
$s_{12}s'_{12}=\lambda_d/\lambda_s |\eps_d a^2|$, where $s'_{12}$ stands
for the ``Cabibbo'' mixing of the right-handed states.
If $s'_{12}<s_{12}\approx(m_d/m_s)^{1/2}$, then this equation
implies $|\eps_d a^2|<1$, so that $M^{\rm max}_t$ becomes smaller.
On the contrary, the upper bound on $M_t$ can be lifted if
$s'_{12}\gg (m_d/m_s)^{1/2}$. However, such an enhancement of
$s'_{12}$  above the naively expected size $O(\eps_d$)
implies substantial fine tuning.

\vspace{5mm}
{\large \bf 3. Discussion and Outlook}
\vspace{2mm}

We have considered the SUSY $SO(10)$ model with small $\tan\beta$,
in which the flavour structure arises in a mass matrix
$\hat{M}_F$ of the superheavy $F$ fermions and is transfered to
the light fermions in an inverse way by means of the universal seesaw.
The largest eigenvalue of $\hat{M}_F$ is given by the $SO(10)$
invariant mass $M\gg M_X$ and thus is unsplit. The lighter eigenvalues,
respectively of the order of $M_X$ and $M_X^2/M$,
arise due to the couplings with the Higgs $45_X$ and are thereby split.

As a result, the quark and lepton Yukawa matrices
have the {\em inverse hierarchy} form (\ref{mf-1}).
The hierarchy of Yukawa constants is described by the approximate
scaling low $\la_{fi}\sim \la\eps_f^{1-i}$, where $i=1,2,3$ is the
family number and the expansion parameters
$\eps_{e,u,d}\!\sim\! M_X/M\!\sim\! 10^{-1}\!-10^{-2}$
are related by the $SO(10)$ symmetry as $\eps_e=-(\eps_d+2\eps_u)$.
As far as $M_X\sim 10^{16}\,$GeV, the above estimate points to the
scale $M\sim 10^{17}-10^{18}\,$GeV, close to the string scale.
The $\la_b\approx \la_b$ unification at the GUT scale implies that
$\eps_u\ll \eps_e \approx -\eps_e$. By this reason, the `up-down'
splitting is quickly growing with the family number:
$\la_c/\la_s\sim |\eps_d/\eps_u|$ and
$\la_t/\la_b\sim |\eps_d/\eps_u|^2$.
The first family plays a role of the {\em Yukawa unification
point} ($\la_{e,d,u}\sim \la$),  with its splitting
understood by the same mechanism that enhances the Cabibbo angle
up to the value $s_{12}\simeq\sqrt{m_d/m_s}$. The other mixing angles
stay much smaller (see (\ref{Cabibbo})).
For the light quarks we have obtained
$m_s\simeq 150$ MeV, $m_d\simeq 7$ MeV and $m_u/m_d\simeq 0.5-0.7$.
The upper limit on the top mass in our scheme is about 165 GeV, which
can be marginally enhanced up to 170 GeV. On the other hand,
the lower bound $M_t\geq 160$ GeV implies $m_s/m_d\geq 22$.
It is worth to mention that small values $\tan\beta=1.4-1.7$ are of
phenomenological interest in testing the MSSM Higgs sector
at new colliders \cite{kun}.

We find it amusing that the inverse hierarchy ansatz implemented in
SUSY $SO(10)$ theory reproduces the fermion mass and mixing pattern
in a very natural and economical way. Our approach is rather general,
with the key assumption that the fermion masses are induced via
seesaw mechanism, by means of the couplings (\ref{WY})
with constants $\hat{Q}_n$ being rank-1 matrices.
We have not specified the concrete symmetry reasons
that could support our ansatz.
Various possibilities can be envisaged, including normal or R-type
discrete and abelian symmetries.
(For example, the combination of such a symmetries fixing the proper
operator structure for the ansatz \cite{DHR} have been found
recently \cite{HR}.)
Notice, that in contrast to the known predictive frameworks
\cite{Fritzsch,DHR} we did not exploit any particular
{\em zero texture}: except that $\hat{Q}$'s are assummed to be
the rank-1 matrices, the Yukawa constants are left completely general.
By this reason, the amount of exact predictions in our scheme is less
than e.g. in \cite{DHR}.
Clearly, a number of free parameters can be reduced by imposing a
proper horizontal symmetry which can restrict the Yukawa matrices
at the needed degree and thus enhance predictivity.
Last but not least, a {\em clever} horizontal symmetry seems to be needed
also for evading a potential problem of too large rates for the lepton
flavour violating processes \cite{BH}, which in our scheme should be
induced due to the presence of large
Yukawa constants above the GUT scale.


\newpage

\thispagestyle{empty}

{\large Figure Caption}

\vspace{7mm}

\vspace{3mm}
Fig. 1. Solid curves show the prediction of our ansatz for
the $\tilde{\la}_t-\tan\beta$ correlation
for given $m_s/m_d$ and $\Ga_{c}$ ($\zeta=22$ and $\Ga_{c}=1.5$, 2.2,
3.3, 6.6 (Fig. A), and $\zeta=25$ and $\Ga_{c}=3.3$ (Fig. B)).
Other input parameters are fixed
as $\alpha_3(M_Z)=0.11$, $m_b=4.4\,$GeV and $m_c=1.32\,$GeV.
Isocurves for fixed top mass are dashed:
$M_t=150$, 160, 170 and 180 (in GeVs).
The isocurves corresponding to different values of $m_u/m_d$ are also
shown: $\rho=0.4$, 0.5, 0.6, 0.7 and 0.8 (dotted).


\begin{thebibliography}{99}


\bibitem{Fritzsch} H. Fritzsch, Nucl. Phys. {\bf B 155} (1979) 189;
H. Georgi and C. Jarlskog, Phys. Lett. {\bf B86} (1979) 297;
J. Harvey, P. Ramond and D.B. Reiss,
Nucl. Phys. {\bf B 199} (1982) 223;
S. Dimopoulos, L.J. Hall and S. Raby, Phys. Rev. Lett. {\bf 68} (1992)
1984;
H. Arason, D.J. Castano, P. Ramond and E.J. Piard, Phys. Rev. D {\bf 47}
(1993) 232;
P. Ramond, R.G. Roberts and G.G. Ross, Nucl. Phys. {\bf B 406} (1993) 19.

\bibitem{DHR} G. Anderson, S. Dimopoulos, L. Hall, S. Raby and G. Starkman,
Phys. Rev. D {\bf 49} (1994) 3660.

\bibitem{tanb} A.E. Nelson and L. Randall, Phys. Lett. {\bf B316} (1993)
516; L.J. Hall, R. Rattazzi and U. Sarid,
preprint SU-ITP-94-15 (1994).

\bibitem{heavyferm} C.D. Frogatt and H.B. Nielsen, Nucl. Phys.
{\bf B147} (1979) 277;
Z.G. Berezhiani, Phys. Lett. {\bf B129} (1983) 99; {\bf B150} (1985) 177;
S. Dimopoulos, Phys. Lett. {\bf B129} (1983) 417;
J. Bagger, S. Dimopoulos, H. Georgi and S. Raby, in Proc Fifth Workshop
on {\em Grand Unification}, eds. K. Kang et al., World Scientific,
Singapore, 1984.


\bibitem{Rattazzi} Z.G. Berezhiani and R. Rattazzi, Nucl. Phys. B {\bf 407}
(1993) 249;
JETP Lett. {\bf 56} (1992) 429;
Phys. Lett. {\bf B279} (1992) 124.

\bibitem{Kazim} For more details of the model, see Z.G. Berezhiani,
Proc. XVI Int. Warsaw Meeting "New Physics with New Experiments",
Eds. Z. Ajduc {\em et al.}, World Scientific, Singapore, 1994, p. 173
[hep-ph/9312222]; Proc. Int. Workshop "SUSY 94", Eds. C. Kolda and
J.D. Wells, UM-TH-94-35, 1994, p. 42 [hep-ph/9407264].

\bibitem{BabuBarr} K.S. Babu and S.M. Barr, Phys. Rev. D {\bf 48}
(1994) 5354.



\bibitem{DiWi} S. Dimopoulos and F. Wilczek, Preprint NSF-ITP-82-07,
1982 (unpublished);
M. Srednicki, Nucl. Phys. {\bf B202} (1982) 327.






\bibitem{Nath} P. Nath, A. Chamseddine and R. Arnowitt, Phys. Rev. D
{\bf 32} (1985) 2348;
R. Arnowitt and P. Nath,
Phys. Rev. D {\bf 49} (1994) 1479.

\bibitem{RG} V. Barger, M. Berger and P. Ohmann, Phys. Rev. D {\bf 47}
(1993) 1093;
S. Naculich, Phys. Rev. D {\bf 48} (1993) 5293.

\bibitem{Shifman} M. Shifman, preprint TPI-MINN-94/42-T (Dec 1994),
 hep-ph/9501222.

\bibitem{Lang} P. Langacker, in Precision Tests of the Standard Model,
World Scientific, Singapore, 1994 [hep-ph/9412361].

\bibitem{contra} M. Consoli and F. Ferroni, Phys. Lett. {\bf B349}
(1995) 375.

\bibitem{Nir} P. Langacker and N. Polonsky, preprint UPR-0642T (March 1995).


\bibitem{inffix} J. Giveon, L. Hall and U. Sarid, Phys. Lett. {\bf B271}
(1991) 138;
M. Carena {\em et al.}, Nucl. Phys. {\bf B379} (1992) 33;
V. Barger {\em et al.},
Phys. Lett. {\bf B314} (1993) 351;
P. Langacker and N. Polonsky, Phys. Rev. D {\bf 47} (1993) 4028; D {\bf 49}
(1994) 1454.

\bibitem{uds} H. Leutwyler, Nucl. Phys. {\bf B337} (1990) 108;
J.F. Donoghue and D. Wyler, Phys. Rev. {\bf D45} (1992) 892.

\bibitem{CDF} CDF Collaboration, F. Abe {\em et al.}, Phys. Rev. Lett.
{\bf 74} (1995) 2626.

\bibitem{inprep} Z. Berezhiani, in preparation.

\bibitem{kun} R. Barbieri, M. Frigeni and F. Caravaglios, Phys. Lett.
{\bf B258} (1991) 167;
Z. Kunszt and F. Zwirner, Nucl. Phys. {\bf B 385} (1992) 3.

\bibitem{HR} L.J. Hall and S. Raby, preprint OHSTPY-HEP-T-94-023
(Jan 1995).

\bibitem{BH} R. Barbieri and L. Hall, Phys. Lett. {\bf B338} (1994) 212;
R. Barbieri, L. Hall and A. Strumia, preprint IFUP-TH 72/94 (Jan 1995).

\end{thebibliography}
\end{document}